# Healthy Lifestyles and Self-Improvement Videos on YouTube:
# A Thematic Analysis of Teen-Targeted Social Media Content

Kyuha Jung, MA[1, *], Tyler Kim[2, *], Yunan Chen, PhD[1],
[1]University of California, Irvine, California, USA;
[2]South High School, Torrance, California, USA

**Abstract**
*As teenagers increasingly turn to social media for health-related information, understanding the values of teen-targeted content has become important. Although videos on healthy lifestyles and self-improvement are gaining popularity on social media platforms like YouTube, little is known about how these videos benefit and engage with teenage viewers. To address this, we conducted a thematic analysis of 44 YouTube videos and 66,901 comments. We found that these videos provide various advice on teenagers' common challenges, use engaging narratives for authenticity, and foster teen-centered communities through comments. However, a few videos also gave misleading advice to adolescents that can be potentially harmful. Based on our findings, we discuss design implications for creating relatable and intriguing social media content for adolescents. Additionally, we suggest ways for social media platforms to promote healthier and safer experiences for teenagers.*

**Introduction**

Along with physical health, adolescence between the ages 13 to 18 is a critical period for mental health development, shaping emotional, cognitive, and social well-being in ways that could have lifelong consequences[1]. Promoting good mental health during this stage of life can help prevent mental health conditions and support teenagers in building resilience, coping with challenges, and thriving in their daily lives[2]. Given its lasting impact, it is essential to provide the right guidance to adolescents that could positively impact their overall health and well-being, such as by encouraging healthy lifestyles and self-improvement.

Among many potential approaches, social media has long been recognized as a helpful, technology-mediated way to support healthy lifestyles and self-improvement. For instance, Instagram has become an online community for sharing healthy habits and mood-tracking experiences[3], where users upload their tracking progress and exchange tips and motivation. Moreover, chronic illness patients use social media platforms like YouTube and TikTok to express their lived experiences through videos and find social support from their viewers[4,5]. On the other hand, harms exist with the use of social media in health contexts, including negative reinforcement of body images[6] and the prevalence of health misinformation for marginalized populations like transgenders[7].

Especially for the adolescent populations, the role of social media in influencing teenagers' health and well-being has gained much public attention[8], and yet the debate remains inconclusive[9]. Previous studies have confirmed that social media platforms can be effective in increasing health knowledge for adolescents[10], whereas Haslam and colleagues[11] raised concerns about the credibility of health-related content on YouTube. Regardless, social media has become a preferred source for adolescents to find various health-related information[12,13], including gaining insights on healthy lifestyles and self-improvement.

Indeed, over one million views from many teen-targeted YouTube videos like "*23 Life Lessons for Teenagers*[14]," "*If you're 13-18 years old, please watch this video...*[15]," or "*how to CONQUER high school*[16]" suggest teenagers' interests in learning about healthy lifestyles and self-improvement. However, less is known about how this content specifically informs and helps its intended teenage audience. This becomes particularly important for institutions like public health agencies, youth organizations, and school districts that strive to use social media for health communication, but often face challenges in making content helpful and engaging for the adolescent population[17].

To address this issue, our study investigates the values of teen-targeted social media content on healthy lifestyles and self-improvement. We explore the following research questions: 1) *How can teen-targeted social media content on healthy lifestyles and self-improvement be useful to teenagers?*, 2) *How can teen-targeted social media content on*

---
[*] The authors contributed equally to this work.

*healthy lifestyles and self-improvement be engaging in delivering health-related information?*, and 3) *How do teenagers respond to teen-targeted social media content on healthy lifestyles and self-improvement?*

We collected and analyzed 44 teen-targeted YouTube videos on healthy lifestyles and self-improvement, with 77% of them (34 videos) surpassing 100,000 views. Through a thematic analysis of the video content and viewer comments, we discovered that creators offered various advice on healthy lifestyles and self-improvement to tackle teenagers' common issues. Advice ranged from mental health and well-being, productivity, social skills, and academic and career success. Additionally, authentic narratives were instrumental in making the videos interesting as creators used informative, reflective, and illustrative storytelling. Viewer comments revealed that viewers formed a teen-centered community, exchanging support and encouraging self-expression. We also found unhealthy advice from a few videos containing potentially harmful suggestions for adolescents. In the Discussion, we provide design implications for creating teen-targeted social media content on health-related topics and suggestions for social media platforms to foster positive experiences for adolescents.

**Methods**
We collected and analyzed videos and their comments from YouTube, a widely used social media platform among adolescents. In this section, we describe our data collection, preprocessing, and analysis process.

*Data Collection and Preprocessing*
We used *yt-dlp*[18], an open-source command-line tool to search YouTube videos and their comments. Following internal pilot tests by KJ and TK, we created 22 search queries by combining 11 adolescent-related keywords ("*teenager*," "*middle school*," "*junior high*," "*high school*," "*13-18 years*," "*13 years*," "*14 years*," "*15 years*," "*16 years*," "*17 years*," "*18 years*") with 2 topic keywords ("*healthy lifestyles*" and "*self-improvement*"). For example, our queries included "*teenager healthy lifestyles*," "*high school self-improvement,*" or "*13-18 years healthy lifestyles*." For each video, we extracted metadata such as title, URL, view count, upload date, creator name, and length in seconds.

From our initial search with *yt-dlp*, we collected 2,900 videos. To refine this dataset (See Figure 1 and Table 1), we applied 3 automatic screening criteria within our data collection script. First, we excluded videos shorter than 3 minutes to focus on longer, more detailed content (C1). Second, we removed videos whose titles did not include explicit teen-related terms such as "*teen*," "*teens*," "*middle school*," "*junior high*," "*high school*," "*13*," "*14*," "*15*," "*16*," "*17*," and "*18*" (C2). Third, we excluded videos with 10,000 views or fewer to focus on videos that had gained a certain level of viewership (C3). Through this process, we eliminated 2,387 videos: 1,756 due to C1, 422 due to C2, and 209 due to C3.

After automatic screening, we noticed that several duplicates existed because identical videos with the same title, view count, and duration had been collected from different queries. To address this, we removed 138 duplicate videos.

Following duplicate removal, we used 2 more manual screening criteria to ensure the remaining videos were appropriate for analysis. First, we watched the videos to confirm that they primarily discussed healthy lifestyles or self-improvement (C4). Even if the video title contained teen-relevant keywords, we removed it if the content focused on less relevant topics such as aesthetic appearances, fashion, or finances. Second, we excluded videos that were not in English or did not provide English captions (C5). This manual screening resulted in the removal of 337 videos: 335 due to C4 and 2 due to C5.

Lastly, to ensure that our dataset was comprehensive, we ran additional rounds of personal searches on YouTube and identified 6 more videos to include. In the end, our final dataset consisted of *44 videos* and *66,901 comments*.

**Figure 1.** Data Collection and Preprocessing Process

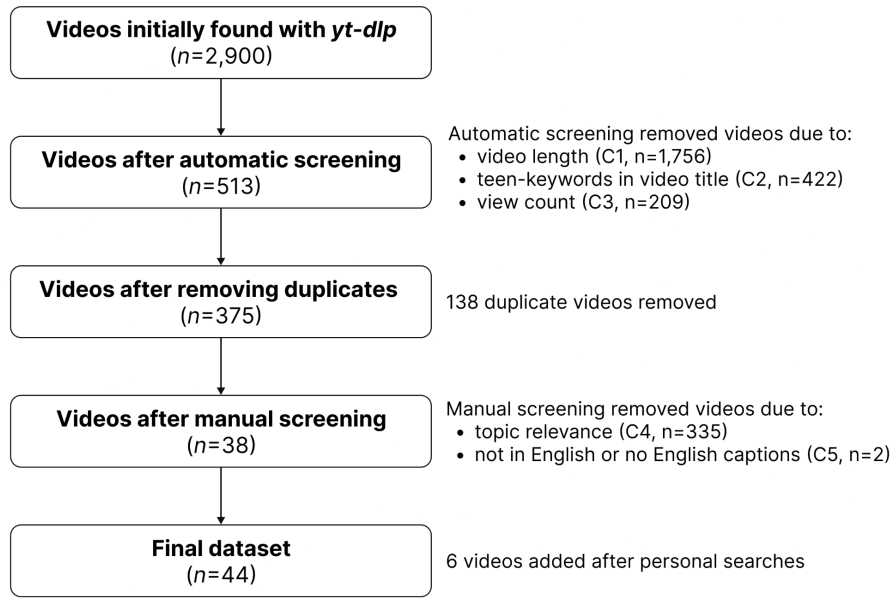

**Table 1.** Inclusion/Exclusion Criteria and Data Preprocessing Methods

|    | **Inclusion Criteria** | **Exclusion Criteria** | **Preprocessing Method** |
|----|------------------------|------------------------|--------------------------|
| C1 | The video is longer than 3 minutes. | The video is short-form content, or exactly or shorter than 3 minutes. | Automatic screening by video length |
| C2 | The video's title includes explicit keywords related to adolescent populations (e.g., "*teen*," "*high school*," or "*13-18 years*"). | The video's title does not include explicit keywords related to adolescent populations (e.g., "*teen*," "*high school*," or "*13-18 years*"). | Automatic screening by video title |
| C3 | The video has more than 10,000 views. | The video has exactly or less than 10,000 views. | Automatic screening by view count |
| C4 | The video primarily discusses topics on healthy lifestyles and self-improvement. | The video does not primarily discusses topics on healthy lifestyles and self-improvement (e.g., aesthetic appearances or fashion). | Manual screening by watching the videos |
| C5 | The video is made in English or provides English captions. | The video is not made in English and does not provide English captions. | Manual screening by watching the videos |

*Data Overview*
Our final dataset was 44 videos created by 35 different creators, 11 of whom had been making content since they were teenagers. The length of these videos (see Table 2) ranged from 257 seconds (4 minutes and 17 seconds) to 4,009 seconds (66 minutes and 49 seconds), with an average length of 804.80 seconds (about 13 minutes and 25 seconds) and a median length of 1,085.50 seconds (about 18 minutes and 6 seconds).

As of March 2025, 14 videos (31.82%) had over 1 million views, 20 videos (45.45%) had between 100,000 and 1 million views, and 10 videos (22.73%) had fewer than 100,000 views. Similarly, 10 creators (28.57%) had over 1 million channel subscribers, 16 creators (45.71%) had between 100,000 and 1 million subscribers, and 9 creators (25.71%) had fewer than 100,000 subscribers. While view counts and subscriber numbers can change over time, we include these figures to demonstrate that most of the analyzed videos and creators had significant viewership and popularity. However, we note that these numbers do not necessarily reflect an exclusively adolescent audience.

The 44 videos received a total of 66,901 comments. The number of comments per video ranged from 14 to 12,000, with an average of 1,555.84 comments and a median of 797 comments.

**Table 2.** Descriptive Statistics of Analyzed Videos

| **Videos** (*n*=44) | | **Statistics** |
|---|---|---|
| **Video Length** | Mean | 804.80 seconds (about 13 minutes and 25 seconds) |
| | Median | 1085.50 seconds (about 18 minutes and 6 seconds) |
| | Min / Max | 257 / 4009 seconds (4 minutes and 17 seconds / 66 minutes and 49 seconds) |
| **View Counts** (as of March 2025) | More than 1 million views | 14 / 44 (31.82%) |
| | 100,000 to 1 million views | 20 / 44 (45.45%) |
| | Less than 100,000 views | 10 / 44 (22.73%) |
| **Creator Channel Subscribers** (as of March 2025) | More than 1 million subscribers | 10 / 35 (28.57%) |
| | 100,000 to 1 million subscribers | 16 / 35 (45.71%) |
| | Less than 100,000 subscribers | 9 / 35 (25.71%) |

*Data Analysis and Ethical Considerations on Publicly Available Data*
We conducted a thematic analysis[19] of the collected videos and comments to explore the values embedded in teen-targeted social media content on healthy lifestyles and self-improvement. TK led the open coding process by watching the videos multiple times and inductively coding their characteristics like content, structure, and viewer reactions into a separate spreadsheet. After each round of open coding, TK discussed the codes with KJ weekly and they rewatched the videos together to ensure a shared understanding. KJ also reviewed progress regularly with YC to exchange feedback on the coding process.

As YouTube comments follow a nested structure (where users can leave comments in response to existing comments), we accessed the comments through the video's URL to better understand these interactions and copied them into the spreadsheet. Examples of our codes included "*mental health advice*," "*friendship advice*," "*personal anecdotes*," "*talking head*," "*gratitude*," and "*encouragement*." After five rounds of open coding and team discussions, we iteratively merged related codes into 3 overarching themes: "*Video Advice*," "*Video Narratives*," and "*Comments*." For instance, codes such as "*avoid negative self-talk*" and "*mindfulness activities*" were grouped under "*Advice on Mental Health and Well-Being*," which was then consolidated into "*Video Advice*" along with other types of advice. Our findings are reported based on these themes.

As YouTube videos and comments are publicly available data, an IRB approval was not required by our university. However, we align with researchers who emphasize ethical practices when using publicly available data[20]. To protect the privacy of creators and users, we anonymized their names, used they/them pronouns, paraphrased quotes, and selected and blurred video screenshots with minimal identifiable information.

*Limitations*
Despite our efforts to maintain methodological rigor, our study has some limitations. First, while *yt-dlp* is a widely used open-source tool for searching YouTube videos, its exact search process and capabilities remain uncertain. Although we supplemented this by conducting additional personal searches on YouTube, we may have missed relevant videos that fit our criteria. Second, our analysis focused on analyzing video content and comments, which may not fully capture the experiences of teenage viewers. Although TK, a teenage viewer and a content creator of healthy lifestyle and self-improvement videos, provided valuable insights, our understanding may not reflect the diversity of adolescent perspectives. Third, healthy lifestyles and self-improvement are broad concepts with varying meanings and contexts. As we used only these two keywords into our search queries, other relevant keywords could have yielded additional content for analysis.

**Findings**
Our findings shed light on how teen-targeted social media content addresses healthy lifestyles and self-improvement to the adolescent audience. First, we discuss the comprehensive advice offered in the videos to help teenagers navigate common life challenges. Next, we highlight how content creators use authentic narratives to engage teenage viewers. We then examine how the comments section functions as a teen-centered community where teenagers express gratitude, seek advice, and share their thoughts. Lastly, we bring up instances where a few videos provided unhealthy advice to adolescents.

*Comprehensive Advice for Managing Teenagers' Common Struggles*
Overall, the videos provided a range of advice to help teenagers manage common teen-specific challenges such as emotional struggles, time management, and peer pressure. They provided practical strategies, new perspectives, and opportunities for self-reflection, making the content highly relevant to adolescent viewers. Each video shared a collection of advice, encouraging teenagers to develop healthier mindset and daily habits for wellness and personal development. From our analysis, we identified four key themes in the provided advice (see Table 3), which addressed teenagers' common life struggles: (1) Mental Health and Well-Being, (2) Productivity and Time Management, (3) Social Skills, and (4) Academic and Career Success.

*Advice on Mental Health and Well-Being.* The most common theme, appearing in 30 videos (68.18%), emphasized the importance of maintaining good mental health and well-being during adolescence. Creators shared advice on understanding and handling emotions, such as viewing failures as growth opportunities (V23, V41), avoiding negative self-talk (V36, V12), and appreciating happy moments (V18, V28). Many videos aimed to validate teenagers' feelings, helping them understand their emotional struggles and build resilience. For instance, V14 reassured teenage viewers by sharing that they and their friends had also found adolescence to be one of the most challenging periods of their lives. By normalizing these experiences and explaining why teenagers might feel overwhelmed, the content creators attempted to help viewers make sense of their emotions and feel less alone.

*Advice on Productivity and Time Management.* The creators also advised on teenagers' issues with productivity and time management, appearing in 28 videos (63.64%). They encouraged teenagers to adopt healthier daily routines and manage their responsibilities effectively. Popular tips included waking up early (V16, V40), planning ahead (V5, V24), exercising regularly (V12, V18), and getting enough sleep (V39, V13). These actionable suggestions focused on helping teenagers struggling with workload and scheduling. Many creators also demonstrated these habits in their videos. For example, V37 featured their daily routine as a high school student, showing how they incorporated morning exercise, cooking and eating healthy food, and tasks like homework and chores into their life. By showcasing real-life examples of managing productivity and daily schedules, the videos strived to deliver advice that feel relatable and achievable for teenage viewers.

*Advice on Social Skills.* Another frequent challenge for teenagers highlighted in the videos was developing social skills and managing relationships. 19 videos (43.18%) concentrated on guiding teenagers to improve their social interactions and better navigate relationships with peers and adults. This advice emphasized the importance of surrounding oneself with supportive friend groups (V2, V32), resisting negative peer pressure (V41), and accepting that friendships naturally change over time (V18, V40). The creators provided reassurance and direction on social interactions, recognizing that teenagers often struggle with peer influence and shifting friendships. By sharing the creators' perspectives and experiences in social situations, their videos aimed to support viewers to feel more confident in managing their social well-being.

*Advice on Academic and Career Success.* Lastly, 6 videos (13.64%) provided help on academic and career success. These videos featured practical study tips and strategies, such as V22 which suggested starting to study early, using active recall for better memorization, and developing a positive attitude toward exams. Some creators also encouraged teenagers to explore career paths early to prevent future regrets (V24) while highlighting soft skills as crucial for long-term professional success (V34). These videos were designed to guide teenagers to improve their academic performance and broaden their outlook beyond school as they think about their future careers.

**Table 3.** Categories of advice provided in the videos. Many videos provided advice from multiple categories.

| Categories | Description and Examples | Appearances in Videos (n=44) |
|---|---|---|
| Mental Health and Well-Being | Advice on managing emotions and fostering resilience (e.g., viewing failures as growth opportunities and avoiding negative self-talk) | 30/44 (68.18%) |
| Productivity and Time-Management | Advice focused on improving daily routines and effectively managing responsibilities (e.g., waking up early, planning ahead, and exercising regularly) | 28/44 (63.64%) |
| Social Skills | Advice on building supportive friendships and handling peer pressure (e.g., surrounding oneself with supportive friends and accepting changing friendships) | 19/44 (43.18%) |
| Academic and Career Success | Advice on improving academic performance and preparing for future careers (e.g., starting to study early, using active recall for memorization, and exploring career paths early) | 6/44 (13.64%) |

***Engaging Video Presentation through Authentic Narratives***
Beyond the advice itself, we discovered that authentic narratives played a crucial part in delivering creators' advice and capturing teenage viewers' attention. As viewers could quickly lose interest in the videos, especially with topics related to healthy lifestyles and self-improvement, creators often used storytelling techniques to make their content more engaging and relatable to adolescents. They tailored their video narratives to the unique needs and experiences of adolescents, cultivating a sense of relatability and authenticity between the creators and the teenage audience. Three common narrative styles emerged: (1) Informative, (2) Reflective, and (3) Illustrative, seeking to create an authentic connection using different strategies.

*Informative Narrative.* An informative narrative used a grounded tone to provide practical advice. Creators often filmed in a talking-head style, speaking directly to the camera as if offering a one-on-one coaching session. The tone was calm and confident, bringing up teenagers' struggles and discussing their thoughts and solutions. An example is V36, as the creator talked about teenagers' challenges in managing self-discipline and friendships and offered recommendations from an adult's perspective. An informative narrative made the creator's advice feel credible and trustworthy, teaching teenage viewers clear guidance on handling their issues.

*Reflective Narrative.* Some videos took a reflective approach by sharing creators' personal stories and mistakes from their teenage years. These reflective narratives aimed to connect with teenagers by disclosing creators' vulnerability during their teenage years and encouraging young viewers to learn from their past mistakes. In contrast to the informative style, these videos were more emotional, often evoking feelings of guilt or regret. For instance, in V19, the creator shared their struggles with substance addiction, arrests, and the negative impact it had on their life. By reflecting on their past, they inspired adolescent viewers to avoid similar pitfalls and make healthier life decisions.

*Illustrative Narrative.* An illustrative narrative was more visual, with creators demonstrating their daily routines to highlight healthy habits and methods for personal growth. This approach frequently appeared in vlogs (video blogs, e.g., V37 in Figure 2), giving teenage viewers a behind-the-scenes look at how the advice played out in everyday settings. Compared to simply talking about productivity or wellness, these videos displayed what healthy lifestyles and self-improvement looked like in practice. In V43, the creator shared their daily routines, including waking up early, working out, preparing healthy meals, and journaling at night. These illustrative videos engaged teenage viewers by visually demonstrating the actions behind the advice, making healthy routines more tangible and motivating them to adopt similar habits.

**Figure 2.** Screenshots of Videos using Different Narratives

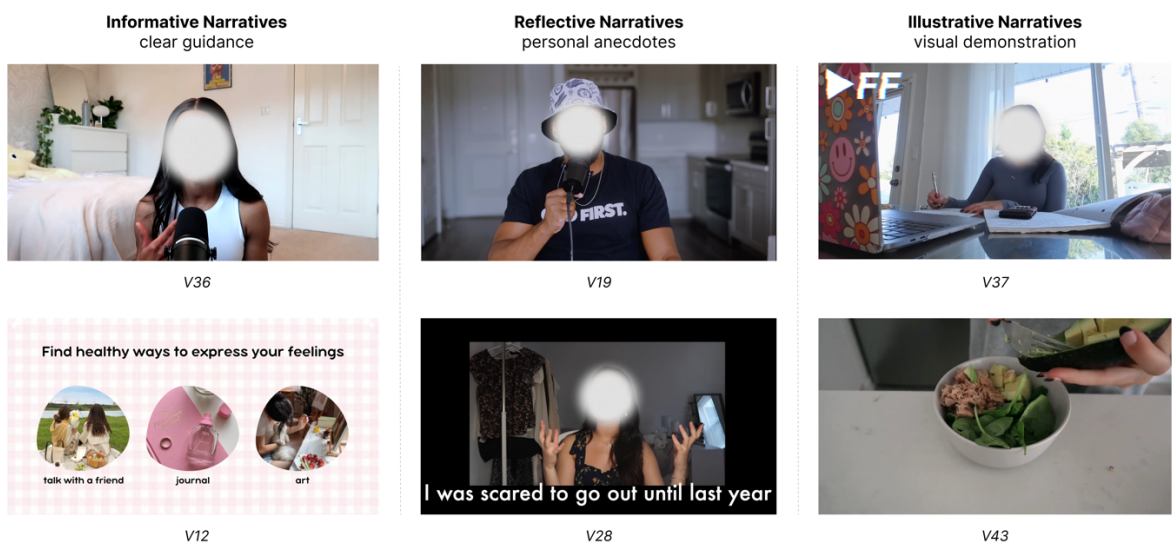

### Emerging Teen-Centered Community Through Comments

In the comments of the videos, viewers created a teen-centered community for them to connect, share, and support one another. While adult viewers occasionally participated by offering advice based on their own experiences (such as in V15, "*I'm 30 years old with two kids, and I just want you to know that you're definitely not alone.*"), it was the teenage audience that predominantly led and shaped the community. Many adolescents engaged by sharing their personal experiences, thoughts, and feelings after watching the videos. One key way they expressed themselves was by disclosing their ages (e.g., "*I'm 15*" or "*I'm in middle school*") in the comments. This served as a means of self-identification and fostered a sense of belonging within the community. In particular, the comment section became a space for teenagers to express gratitude and connection, seek advice, and vent about their life challenges.

First, many teenage viewers used comments to express gratitude and connection. For example, a viewer in V28 shared: "*I've been on antidepressants since I was 10, and now that I'm 16, I'm finally starting to focus on getting better after the trauma caused by my classmates. [...] I just want to say thank you for helping me feel better.*" This type of comment helped to reinforce the creator's message, showing how it resonated with adolescent viewers and inspired them with hope. Furthermore, teenage viewers connected with other viewers of the same age, as seen in the comment in V11: "*I'm 16, and this is probably one of the most helpful videos I've come across lately.*" In response, other 16-year-olds commented, saying, "*I'm 16, too. Where are you from?*", "*Same here,*" and "✋ *[a high-five emoji],*" building a friendly bond among adolescents.

Additionally, adolescents left comments to seek and offer advice. As an example, one viewer in V39 asked: "*I'm 13. Should I spend my time enjoying life and partying, or should I follow the advice you've shared?*" Another viewer, stating their age as a few years older, replied: "*You're 13, bro. Don't waste time partying. You can enjoy life but focus on following the tips on weekdays. Stay happy and keep up the grind.*" Through such exchanges of personalized guidance, teenagers not only received advice from creators but also each other, strengthening a supportive community of peers.

Lastly, the comments often served as an emotional outlet for teenagers, where they vented about their life struggles in detail. For instance, one 16-year-old in V34 elaborated: "*Half a year ago my life turned upside down because I moved to another country. [...] I'm struggling with overweight and anxiety. [...] I am sorry for this long post, but I just needed to write my thoughts somewhere.*" While these comments did not ask for specific advice, they helped teenagers release their emotions through self-disclosure and gain validation from others. In turn, fellow teenage viewers responded with comforting words and shared experiences, such as one 14-year-old who wrote: "*It is not your fault that something like this is happening to you, and you are not talking nonsense. [...] I'm also struggling, but I'm here to tell you that there's always hope.*" As a teen-centered community, the comments became a channel for teenagers to express their inner thoughts and feel heard and less alone in their struggles.

### *Unhealthy Advice in Teen-Targeted Content*
While most videos offered helpful and motivating advice, a few videos (3 videos) promoted potentially biased and unhealthy messages to adolescents. These videos framed their content as self-improvement guidance, but the advice risked being misinterpreted or misused by teenage viewers. For instance, in V26, the creator advised darkening one's beard, taking online business courses, working as an assistant to a social media influencer, and pursuing relationships with women in their mid-20s from stable two-parent households. Additionally, the creator recommended blocking dihydrotestosterone (DHT) to prevent hair loss and linked a hair nutrition product, blending self-improvement advice with product promotion. Similarly, V30's creator encouraged muscle building through intermittent fasting and clean bulking while discouraging relationships with women until viewers had "*grown into a man.*" These videos promoted a narrow and idealized version of masculinity, presenting physical appearance and certain behaviors as the end goal of self-improvement. Many teenage viewers in the comments expressed admiration and agreement, highlighting the influence of such content despite its potentially misleading or harmful nature.

### Discussion
In this section, we discuss our findings in relation to previous research and offer recommendations for designing effective teen-targeted social media content on health-related topics. Specifically, these guidelines may assist public health agencies, youth organizations, and school districts to improve their health communication efforts on social media. We also provide suggestions for social media platforms to create positive experiences for adolescent viewers with content related to healthy lifestyles and self-improvement.

Prior research highlights the importance of relatability in social media content, such as videos that resonate with the target audience's ethno-cultural background[21] or health conditions[4]. Our findings align with this, showing that videos addressing common teenage struggles were especially effective in engaging adolescent viewers with healthy lifestyles and self-improvement. While many creators were adults, discussing issues relevant to teenagers helped capture their attention and provide practical advice. The most frequent topics included mental health and well-being, as well as productivity and time management. As such, social media content focused on teens should be relevant to their lives and needs, incorporating topics like mental health and productivity. Authentic storytelling, like the narratives used by the content creators, could also be used to establish credibility and interests from teen viewers.

In addition to informational support, our findings highlight that teenagers sought emotional support and validation of their challenges. Teenagers found comfort in realizing that they were not alone in their struggles, whether through reassuring messages from creators or supportive comments from other viewers. Previous studies have shown that teenagers use social media to connect with other teens facing similar issues, especially in sensitive contexts like sexual health[22,23]. Given this, teen-targeted content in health and wellness should include emotionally supportive language that acknowledges adolescents' unique hardships and makes them feel understood. Furthermore, social media platforms should foster supportive environments by enabling peer support networks and connecting viewers with similar backgrounds and experiences.

At the same time, we identified instances where potentially harmful advice was presented as healthy lifestyles or self-improvement strategies. In one case, the creator leveraged their trust with teenage audiences by recommending products in their video page. As social media becomes more personalized through algorithmic curation, users may continuously encounter harmful content regardless of their preferences[24]. While online safety policies around adolescent protection are evolving (e.g., Instagram's Teen Accounts[25] introduced in September 2024), it is essential to advocate for strong safeguards on social media platforms. Nevertheless, social media must also respect adolescent users' growing needs for independence and privacy[26] and adopt positive reinforcement methods rather than overly restrictive or intrusive solutions[9]. One approach could be using positive peer dynamics, as our findings revealed that comments from peers could offer both the benefits of forming supportive communities and the risks of normalizing unhealthy ideas. As adolescents could struggle to discern harmful information, especially with peer pressure, social media platforms should implement features that could guide and empower teens in healthier and safer directions.

**Conclusion**
In this study, we explored the values of adolescent-focused social media content on healthy lifestyles and self-improvement. We found that these videos offer diverse advice on common teenage struggles and use engaging narratives to capture viewers' attention. Comments revealed a teen-led community where adolescents could express themselves and seek advice. However, we discovered that a few videos provided harmful advice to adolescents. Our findings provide design implications for creating relatable and intriguing content, encouraging youth health and education organizations to incorporate these insights. We also urge social media platforms to promote safer and healthier experiences for teenagers.